\documentclass[pra,aps,superscriptaddress,twocolumn,showpacs,longbibliography,amsmath,amssymb]{revtex4-2}

\usepackage{ulem}
\usepackage{graphicx}
\usepackage{amsfonts}
\usepackage{color}
\usepackage[dvipsnames,svgnames]{xcolor} 

\usepackage{tikz}
\newcommand*\circled[1]{\tikz[baseline=(char.base)]{
    \node[shape=circle,draw,inner sep=1pt] (char) {#1};}}

\newcommand{\ud}{\mathrm{d}}
\newcommand{\de}[2]{\frac{\ud #1}{\ud #2}}

\newcommand{\pdern}[3]{\frac{\partial^#3#1}{\partial#2^#3}}
\newcommand{\pder}[2]{\frac{\partial#1}{\partial#2}}

\DeclareMathOperator\Arg{Arg}
\usepackage[breaklinks=true]{hyperref}
\graphicspath{ {./figures/} {./}}

\renewcommand{\selectlanguage}[1]{}

\begin{document}

\title{Gain-controlled taming of recurrent modulation instability\\
}

\author{Guillaume Vanderhaegen}
\affiliation{University of Lille, CNRS, UMR 8523-PhLAM-Physique des Lasers Atomes et Mol\'ecules, F-59000 Lille, France}
\author{Pascal Szriftgiser}
\affiliation{University of Lille, CNRS, UMR 8523-PhLAM-Physique des Lasers Atomes et Mol\'ecules, F-59000 Lille, France}
\author{Alexandre Kudlinski}
\affiliation{University of Lille, CNRS, UMR 8523-PhLAM-Physique des Lasers Atomes et Mol\'ecules, F-59000 Lille, France}
\author{Andrea Armaroli}
\affiliation{University of Lille, CNRS, UMR 8523-PhLAM-Physique des Lasers Atomes et Mol\'ecules, F-59000 Lille, France}
\affiliation{Department of Engineering, University of Ferrara, 44122 Ferrara, Italy}
\author{Matteo Conforti}
\affiliation{University of Lille, CNRS, UMR 8523-PhLAM-Physique des Lasers Atomes et Mol\'ecules, F-59000 Lille, France}
\author{Arnaud Mussot}
\affiliation{University of Lille, CNRS, UMR 8523-PhLAM-Physique des Lasers Atomes et Mol\'ecules, F-59000 Lille, France}
\author{Stefano Trillo} 
\affiliation{Department of Engineering, University of Ferrara, 44122 Ferrara, Italy}

\begin{abstract}
We show how the recurrence phenomenon characteristic of the nonlinear stage of induced modulational instability in a passive fiber is  affected by forcing. 
An additional linear amplification, even if extremely weak, induces separatrix crossing in correspondence of critical values of the gain around which the recurrence process considerably slows down, switching between dynamical orbits of different kind.  
We present evidence for such phenomenon in a fiber optics experiment where the gain is finely tuned by means of Raman amplification. A theoretical explanation is also provided that matches almost perfectly with our experimental results.
\end{abstract} 

\date{\today} 
\maketitle
\section{Introduction}

Modulational or Benjamin-Feir instability (MI) \cite{Benjamin_1967, Tai_1986, Zakharov_2009} is, in its most general meaning, the exponential amplification of a small perturbation at the expense of a strong background (pump) wave. In the last decade a lot of attention was devoted to the fully nonlinear stage of MI past the initial amplification. Different scenarios are possible which range from the formation of breather pairs \cite{PhysRevX.5.041026} or the onset self-modulated structures for localized perturbations \cite{PhysRevLett.116.043902,Conforti:18,PhysRevLett.122.054101}, to the recurrence (cycles of conversion and back-conversion between the perturbation and the background) characteristic of purely periodic perturbations (\textit{e.g.}, the typical case of a single injected  sideband pair). The observation of the latter regime, pioneered in hydrodynamics \cite{Lake_1977}, and investigated recently in more depth in optical fibers \cite{Van_Simaeys_2001,Mussot_2018}, and in bulk crystals \cite{Pierangeli_2018}, is particularly intriguing, not only because it is evocative of the famous Fermi-Pasta-Ulam-Tsingou phenomenon \cite{FPUToriginal}, but also because the recurrences are well organized according to a non-trivial phase-space structure ruled by the underlying simplest model, namely the nonlinear Schr\"odinger equation (NLSE). The integrable nature of the NLSE allows for a detailed description of the phase-space associated to MI \cite{Akhmediev_1986,Trillo_1991}, whose main trait is the coexistence of two distinct types of recurrences (denoted below as shifted and unshifted, or resp.~period-2---P2, and period-1---P1), which can be selectively accessed by acting on the launching conditions. 
In particular, the dynamics of nonlinear MI  in optical fibers was shown to be in excellent agreement with the NLSE solutions not only in early demonstrations based on cutback experiments \cite{Van_Simaeys_2001,Hammani_2011,Hu_2018,Bendahmane:15}, 
but also in more advanced results obtained through non-destructive reconstruction of the Fourier mode evolutions in amplitude and phase, performed via heterodyne detection of backscattered light \cite{Mussot_2018,Hu_2018,Naveau_2019,Naveau_2019_2,Naveau_2021,Vanderhaegen_2020_Opex,Vanderhaegen_2020_OL}. 

One of the strongest conditions to observe the exact and rich regular dynamics associated to the NLSE over several recurrence periods, besides the low background noise \cite{Vanderhaegen_2022_therm},  is the conservation of the total field intensity, which is achieved in the recent advanced experiments through counter-propagating Raman amplification. Albeit small, the natural energy dissipation in any propagation medium irreversibly spoils the possibility to observe long-term recurrences of P1 type, as reported in hydrodynamics \cite{Kimmoun_2016}. In this respect, optical fibers turned out the ideal platform to detect the separatrix-crossing phenomenon from P1 to P2 orbits induced by dissipation \cite{Vanderhaegen_2022_loss}. The careful control of  net losses through undercompensated Raman amplification allowed to prove multiple separatrix crossing events occurring at critical values of the loss coefficient. 

This paper is aimed at investigating the analogous phenomenon in the presence of a net gain instead of dissipation. Indeed, in hydrodynamics, forcing naturally occurs from the action of the wind, and this was shown to yield the opposite effect (\textit{i.e.}, crossing from P2 to P1) \cite{Armaroli_2018,Eeltink_2020}. However, an experimental demonstration of the phenomenon can be foreseen with much better precision in optics,
where unprecedented level of control of the forcing can be obtained by tuning the Raman pump to achieve slight overcompensation of the losses. This is the focus of the present work, where we present the first evidence of separatrix crossing induced by the weak linear amplification in optical fibers. 
The possibility to tame the recurrence phenomenon is offered by the dramatic slowing down of the dynamics which is found to occur around the critical values of gain responsible for crossing.

After recalling and adapting to gain the theoretical approach of Ref.~\cite{Vanderhaegen_2022_loss}, which, we recall is based on two different perturbation approaches, we present the physical mechanism for gain tuning and recall the experimental setup. Compared with the case of dissipation, a different trade-off in fiber parameters is required to limit spurious noise amplification.  We then report on the experimental results and their comparison to the theoretical and numerical predictions.

\section{Theory of separatrix crossing from forcing}
Let us consider a generalized nonlinear Schrödinger equation (NLSE) including forcing \cite{Agrawal_2012}
\begin{equation}
    i \frac{\partial E}{\partial Z}-\frac{\beta_2}{2}\frac{\partial^2 E}{\partial T^2}+\gamma|E|^2 E= i\frac{g}{2}E
    \label{eq:NLSE_gain}
\end{equation}
where $\beta_2$ [ps$^2/$km] denotes the group-velocity dispersion (GVD), $\gamma$ [$\mathrm{W^{-1}km^{-1}}$] the nonlinear coefficient, $g$ the linear gain [$\mathrm{km}^{-1}$], and $E$ [$\mathrm{W}^\frac{1}{2}$] the complex envelope of the electric field propagating in the fiber; $Z$ [km] and $T$ [ps] are propagation distance and time in the frame moving at the group velocity.
In the focusing regime ($\beta_2\gamma<0$), the $T$-independent solution of Eq.~\eqref{eq:NLSE_gain} $E=\sqrt{P_p}\exp{(i P_p Z)}$ is modulationally unstable. 

It is convenient to cast Eq.~\eqref{eq:NLSE_gain} in dimensionless form by defining $t=T/T_0$, $z = Z/L_\mathrm{nl}$, $\psi = E/\sqrt{P_\mathrm{tot}}$, with $P_\mathrm{tot}\neq P_p$ the total peak power injected in the fiber---possibly including additional sideband contributions, see below---$L_\mathrm{nl} \equiv \left(\gamma P_\mathrm{tot}\right)^{-1}$, $T_0\equiv \sqrt{L_\mathrm{nl}|\beta_2|}$; finally, denoting $\tilde g \equiv g L_\mathrm{nl}$, we write
\begin{equation}
    i\pder {\psi}{z }+ \frac{1}{2}\pdern{\psi}{t}{2}+\gamma|\psi|^2 \psi = i\frac{\tilde g}{2}\psi 
    \label{eq:NLSEadim_gain}
\end{equation}
 
 A simple, yet qualitatively meaningful, analytical approach based on three-wave mixing (3WM) was obtained in Ref.~\cite{Trillo_1991, Cappellini_1991} for $\tilde g=0$ and used in \cite{Vanderhaegen_2022_loss} to analyze the effect of damping on separatrix crossing. We succinctly adapt it to the NLSE with amplification [Eq. \eqref{eq:NLSEadim_gain}].

We recall that, for $\tilde g=0$, introducing the dimensionless angular frequency $\Omega=2\pi f_m T_0$ related to the real-world modulation frequency $f_m$, MI occurs for $|\Omega| \le 2$. In the simplest case of a single unstable mode $1<\Omega \le 2$, we can consider a perturbing pair of symmetric sidebands, also denoted as signal-idler pair in the following. By defining the following \textit{Ansatz}
\begin{equation} \label{eq:3Wansatz}
\psi(z,t)=  \left[ a_0(z) + \frac{a_1(z)}{\sqrt{2}}\left(e^{i \Omega t} + e^{-i \Omega t} \right) \right] e^{\frac{\tilde g}{2}z}
\end{equation}
and insert it in Eq.~\eqref{eq:NLSEadim_gain}. The variables $a_{0,1}$ represent the complex amplitudes of the carrier and sidebands, respectively. By construction, they satisfy $|a_0|^2+|a_1|^2=1$.

By neglecting all terms oscillating at $\pm m \Omega$, with $m\ge 2$, it is easy to verify that the dynamics takes place in a phase-space defined by two effective conjugate variables: $\Delta\Phi = \Arg\left[a_1\right]-\Arg\left[a_0\right]$ and $\eta=|a_1|^2$. The evolution is conveniently expressed in terms of the rescaled distance $\bar z\equiv \frac{\exp(\tilde g z)-1}{\tilde g}$ as
\begin{equation}
\begin{aligned}
  \de{\eta}{\bar z}&= \frac{\partial H}{\partial \Delta \Phi}; \;\; \de{\Delta\Phi}{\bar z}=-\frac{\partial H}{\partial \eta},\\
 H & = \eta (1-\eta) \cos 2 \Delta\Phi + \left[1-\frac{\Omega^2}{2(1+\tilde g \bar z)} \right] \eta -\frac{3}{4} \eta^2. 
\end{aligned}
\label{eq:H3}
\end{equation}

\begin{figure*}
    \centering
    \includegraphics[width=\textwidth]{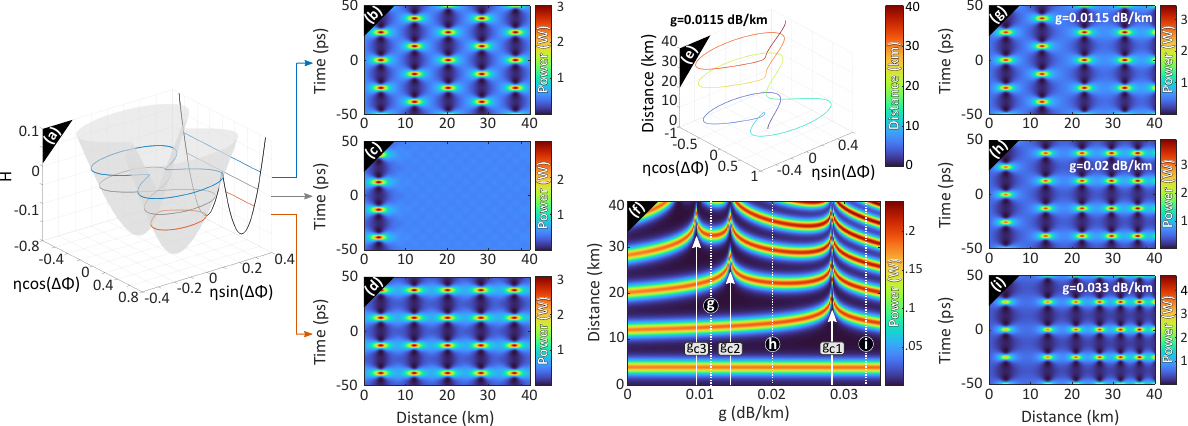}
    \caption{(a) Hamiltonian surface $H=H(x,y)$ of integrable 3WM in the phase-space $(x,y) \equiv (\eta \cos(\Delta \Phi),\eta \sin(\Delta \Phi))$ and its projection on the plane $\Delta \Phi=0$ (black line in the vertical plane behind the $H$-surface), with highlighted level curves corresponding to P2 ($H=-0.05$, orange line) and P1 ($H=0.05$, blue line) orbits around the separatrix (AB, $H=0$, grey line); (b-d) Corresponding space-time power evolutions from the conservative NLSE ($g=0$). (e) Phase-space trajectory vs. distance $Z$ showing separatrix crossing for $g=0.0115$ dB/km). (f) False-color plot of the signal power $P_s$ evolutions along the fiber distance as a function of weak forcing $g$ for initial condition $\Delta \Phi (z=0)=-\frac{\pi}{2}$ initially exciting a P2 orbit. (g-i) Examples of space-time power evolutions for $g=\{0.0115,0.02,0.033\}$ dB/km, corresponding to dashed vertical lines in (f). Parameters: $\beta_2=-21$ ps$^2$km$^{-1}$, $\gamma=1.3$ W$^{-1}$km$^{-1}$, $P_p(z=0)=500$ mW, $P_s(z=0)=P_i(z=0)=5$ mW and $f_m=39.6$ GHz.
    }
    \label{fig:well_gain_breaking}
\end{figure*}

 If $g=0$, Eq.~\eqref {eq:H3} describes a 1 d.o.f.~integrable Hamiltonian system, with $H$ a constant of motion. As discussed in details in \cite{Trillo_1991,Cappellini_1991}, the system evolves like a point mass in a double-well potential, see Fig.~\ref{fig:well_gain_breaking}(a). 
Two distinct types of closed orbits (recurrences) are observed, depending on the initial conditions. 
They are illustrated in Fig.~\ref{fig:well_gain_breaking}(b,d) using dimensional units in order to facilitate the comparison with the experiment. 
These parameters are used: $\beta_2=-21$ ps$^2$km$^{-1}$, $\gamma=1.3$ W$^{-1}$km$^{-1}$, $P_p(z=0)=500$ mW, $P_s(z=0)=P_i(z=0)=5$ mW and $f_m=39.6$ GHz, corresponding to the maximum of MI gain.
For an initial condition with $H>0$ the orbit resembles a figure of eight and corresponds to successive peak conversions which are shifted by half a temporal period or $\Delta \Phi=\pi$ (Fig.~\ref{fig:well_gain_breaking}b), whereas $H< 0$ yields single-well type of orbits with conversion peaks always occurring in phase (Fig.~\ref{fig:well_gain_breaking}d).
We denote the former type of orbits as shifted or P2, and the latter ones as unshifted or P1. They are also denoted as type A or type B with reference to exact doubly-periodic solutions of the NLSE \cite{Vanderhaegen_2020_OL}.
The limit case $H=0$ corresponds to the separatrix, where the orbit decays asymptotically to the background after a single growth stage as shown in Fig.~\ref{fig:well_gain_breaking}(c). In turn this trajectory stands for a portion of the full Akhmediev breather (AB) that connects the background to itself at $z=\pm \infty$ \cite{Akhmediev_1986}. For sufficiently weak input modulation ($\eta(0) \ll 1$), this orbit requires a precise (frequency dependent) input phase $\Delta\Phi_\mathrm{{AB}}=\frac{1}{2} \cos^{-1}(\Omega^2/2-1)$ \cite{Mussot_2018}.
We also emphasize that, though Eq.~\eqref{eq:H3} provides only a qualitatively accurate description of the NLSE dynamics, the phase-space representation remains a very powerful tool suitably valid for projecting exact multi-frequency solutions of the NLSE, too \cite{Vanderhaegen_2020_OL}.

Obviously, when $\tilde g=0$, the conservation of $H$ constrains the system to stay on the orbit dictated by the initial condition, thus preventing the possibility for one type of orbit to cross into the other (different) type. Such possibility, however, becomes allowed for $\tilde g \neq 0$, where the system is no longer structurally stable.
This was experimentally demonstrated by studying the influence of damping ($\tilde g<0$) in fiber optics \cite{Vanderhaegen_2022_loss} {and water waves \cite{Kimmoun_2016}} or forcing ($\tilde g>0$) in water waves \cite{Eeltink_2020}. The former exhibits a transition from P1 to P2 orbits,  vice versa (P2 to P1) the latter. 
Here we present a detailed analysis of the impact of forcing in fiber optics. An illustration of this phenomenon is provided in Fig.~\ref{fig:well_gain_breaking}(e), where the phase-space is stroboscopically expanded in the propagation direction (vertical axis). It is apparent that a trajectory starting from a small $\eta(0) = \eta_0\ll 1$ and $\Delta\Phi(0)=\Delta\Phi_0 = \pm\frac{\pi}{2}$, drops, after two phase-shifted recurrence cycles, onto an unshifted orbit. 

A convenient way to illustrate the overall impact of forcing is to plot the evolution of the signal wave power $P_\mathrm{S}=\frac{\eta P_\mathrm{tot}}{2}$ against distance as a function of $g$. 
In Fig.~\ref{fig:well_gain_breaking}(f) we show the incidence of the linear amplification on an initially shifted excitation [$\Delta \Phi(z=0)=-\frac{\pi}{2}$]. 
It is immediately clear that there exist some critical gain values $g_{\mathrm{c}n}$, $n=1,2,3,\dots$, ordered so that $g_{\mathrm{c}1}>g_{\mathrm{c}2} > g_{\mathrm{c}3} > \ldots$  at which the recurrence process slows down dramatically and yields anomalously large, yet finite (as verified numerically), recurrence distances. 
Such critical points originate from the separatrix crossing during the propagation and delimit transitions from shifted to unshifted recurrences. From the space-time power profiles in Fig.~\ref{fig:well_gain_breaking}(g-i), we deduce a general behavior according to the value of the gain. If $g_{cn}<g<g_{c(n-1)}$, the first $n$ maximum compression points are consecutively shifted by half a temporal period while the successive ones are in-phase with the $n$-{th} one. The separatrix crossing occurs then before---and very close to---the end of the $n$-th passage close to the origin. We recall that, in the case of forcing, the system is not integrable anymore and the recurrence distance is not uniform and can deviate substantially from the doubly-periodic solutions of the NLSE \cite{Conforti_2020, Vanderhaegen_2020_OL}.


To describe such complex dynamics and predict the appearance of the critical gain values for separatrix crossing, we can rely on Eq.~\eqref{eq:H3}, similarly to 
Ref.~\cite{Vanderhaegen_2022_loss}
The critical gain values $\tilde g_{cn}$ for crossing are estimated as the solutions of 
\begin{equation}
    H_{0}=-\frac{\tilde g_{cn}\Omega^2}{2} \int_{0}^{n z_\mathrm{per}} e^{-\tilde g_{cn}z}\eta(z)dz,
    \label{eq:gcrit_3W}
\end{equation}
where $H_0=\eta_0(1-\eta_0)\cos(2\Delta \Phi_0)+(1-\frac{\Omega^2}{2})\eta_0 -\frac{3}{4}\eta_0^2$ is the initial 3WM Hamiltonian,
$z_\mathrm{per}$ is the longitudinal period without forcing, and $\eta(z)$ is the modulation power fraction along the orbit (see Appendix A). 
Indeed, Eq.~\eqref{eq:gcrit_3W} entails that the transition between the P2 and the P1 orbit that occurs when the Hamiltonian crosses the value $H=0$, happens to take place at the $n$-th back-conversion to the pump which in turn corresponds to the $n$-th passage near the origin \cite{Vanderhaegen_2022_loss}. 

\begin{figure}[h!]
    \centering
    \includegraphics[width=\columnwidth]{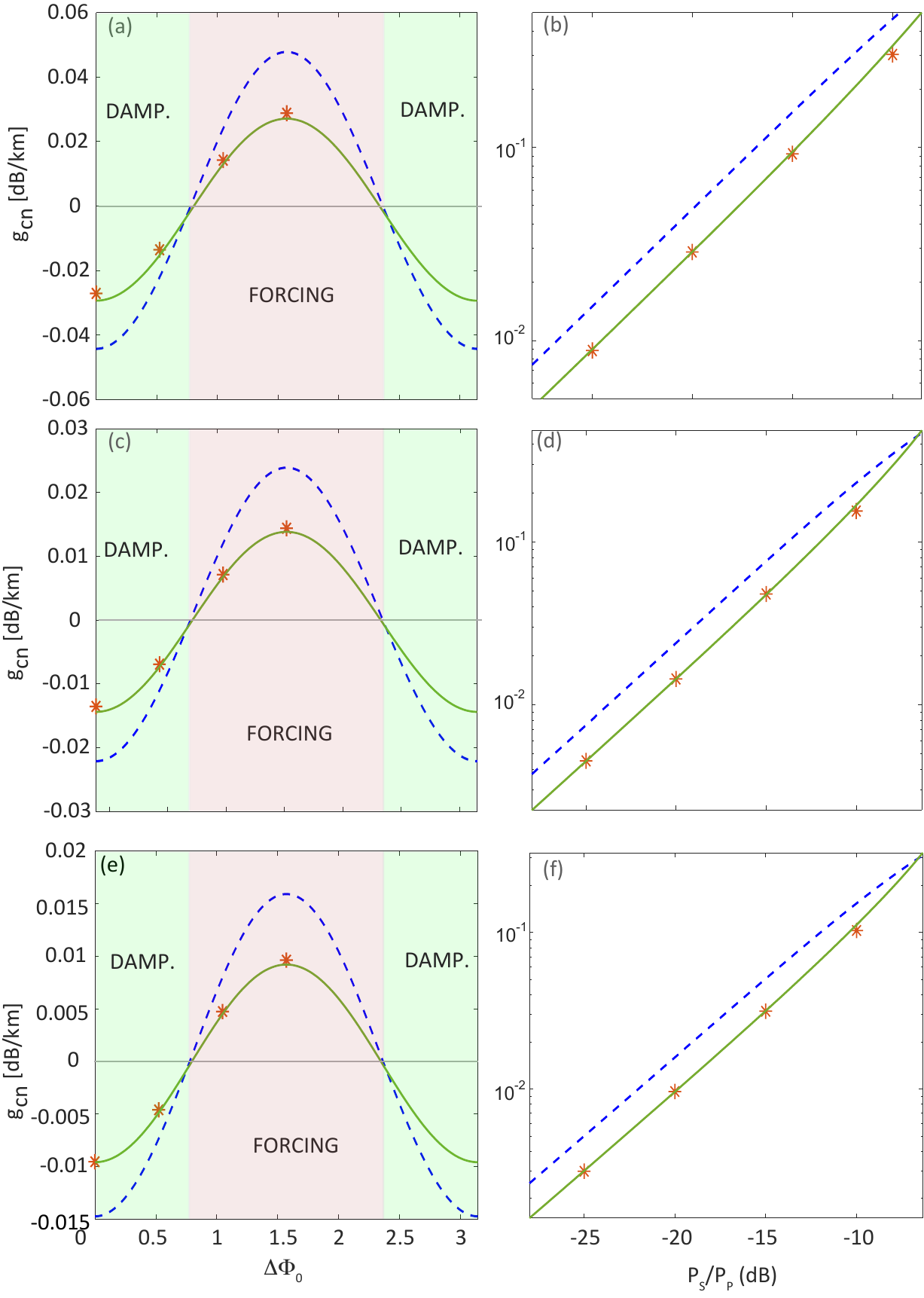}
    \caption{Comparison of critical gain $g_{cn}$ as obtained from 3WM model (Eq. \eqref{eq:gcrit_3W}, dashed blue line), finite-gap approach (Eq. \eqref{eq:finite_gap_gain}, solid green line),  and NLSE simulations (crosses) vs. relative phase $\Delta \Phi_0$ for fixed $\eta_0=0.0196$ (left column panels a,c,e), or vs. sideband to pump power ratio $\eta_0$ for fixed $\Delta \Phi(0)=\frac{\pi}{2}$. The results are relative to first three crossing: $n=1$ (a,b), $n=2$ (c,d), $n=3$ (e,f).
    The other parameters are the same as in Fig.~\ref{fig:well_gain_breaking}. }
    \label{fig:finite_gap_3wm}
\end{figure}

The $\tilde g_{cn}$ values obtained from this Hamiltonian approach of the 3WM model are plotted as a function of the initial relative phase $\Delta\Phi_0$ and the signal to pump ratio $\frac{P_\mathrm{s}}{P_\mathrm{p}}= \frac{\eta_0}{2(1-\eta_0)}$ for $n=\{1,2,3\}$ in Fig.~\ref{fig:finite_gap_3wm} (blue dashed lines) . 
The critical gain $g_{cn}$ as a function of $\Delta \Phi_0$, as shown in Fig.~\ref{fig:finite_gap_3wm}(a,c,e), exhibits a cosine-like shape (symmetric for $\Delta \Phi_0 \rightarrow -\Delta \Phi_0$). There exist a critical value of phase, which obviously corresponds to the separatrix phase $\Delta\Phi_\mathrm{{AB}}$,
at which $g_{cn}=0$. For phase values in $[0;\Delta \Phi_{AB}]$ and $[\pi-\Delta \Phi_{AB};\pi]$, $g_{cn}<0$ (damping), as separatrix crossing occurs from P1 to P2 orbits, as in Ref.~\cite{Vanderhaegen_2022_loss}. For values in $[\Delta \Phi_\mathrm{AB};\pi-\Delta \Phi_{AB}]$, $g_{cn}>0$, \textit{i.e.}, the case we are interested in here (Fig.~\ref{fig:well_gain_breaking}). 
The critical gain values decrease in magnitude as $\Delta\Phi_\mathrm{AB}$ is approached.

Noteworthy, the comparison of 3WM results (dash blue line) and integration of Eq.~\ref{eq:NLSE_gain} (red crosses) shows a discrepancy, which is mainly due to the intrinsic error in the value of {the spatial period} $z_\mathrm{per}$ [see Eq.~\eqref{eq:3Wperiod}, in the Appendix]  compared with the corresponding NLSE estimate.  
A more accurate approach is found in \cite{Coppini_2020}, based on the perturbation theory of finite-gap integration of the NLSE. This approach is valid for small input modulation amplitudes and weak amplification ($\tilde g \ll 1$). From this theory, a simple formula can be derived for the critical gain values $\tilde g=\tilde g_{cn}$ (for details we refer the reader to \cite{Coppini_2020,Vanderhaegen_2022_loss}):
\begin{equation}
    \tilde g_{cn}=\frac{\eta_0 (1-  \eta_0) e_{+} e_{-}}{2ng_\mathrm{MI}},
    \label{eq:finite_gap_gain}
\end{equation}
where $\eta_0$ is the modulation power fraction, 
and $e_{\pm}\equiv \frac{1}{\sqrt{2}} \left(e^{i\mp \psi} e^{-i\Delta\Phi_0} - e^{i\pm \psi} e^{i\Delta\Phi_0}\right) $, with $\psi\equiv \cos^{-1} \frac{\Omega}{2\sqrt{1-\eta_0}}$ are the growing and decaying eigenvectors of MI, as defined in \cite{Grinevich_Santini_2018,Coppini_2020} and $g_\mathrm{MI}=\frac{\Omega}{2} \sqrt{4(1-\eta_0)-\Omega^2}$ is the normalized MI gain. 

We notice in Fig.~\ref{fig:finite_gap_3wm}(a,c,e) that this analytical estimate (green solid lines) follows the same cosine trend in $\Delta \Phi (z=0)$ as for the 3WM model, but with an error below 5$\%$, compared to nearly a factor of two overestimate for the 3WM at $\Delta\Phi_0=\frac{\pi}{2}$. 

It is also interesting to consider how $g_{cn}$ depend on $\frac{P_\mathrm{S}}{P_\mathrm{P}}$ (or equivalently $\eta_0$). We show the numerical as well as the analytical results in Fig.~\ref{fig:finite_gap_3wm}(b,d,f), for $n=1,2,3$. The critical value grows exponentially with the signal power fraction. Both analytical estimates are qualitatively satisfactory. While Eq.~\eqref{eq:gcrit_3W} always grossly overestimates $g_{cn}$, Eq.~\eqref{eq:finite_gap_gain} is almost indistinguishable from numerical results at small $\frac{P_\mathrm{S}}{P_\mathrm{P}}<-15$ dB, while it overestimates $g_{cn}$ at larger ratios; the error at $\frac{P_\mathrm{S}}{P_\mathrm{P}}<-10$ dB is less than $15\%$, though. This can be ascribed to the main assumption of the finite-gap approximation, \textit{i.e.}, a weak initial modulation.

After having assessed the validity and limits of our analytical estimates, we describe the experimental setup and compare numerical and analytical results to experimental recordings.


\section{Experiment}
\subsection{Experimental setup}

To experimentally study  separatrix crossing in the nonlinear stage of MI induced by a weak forcing, the key feature is the careful control of the net gain experienced by the signal during the propagation in the fiber. To this end, we use the backward Raman pumping scheme. While in \cite{Mussot_2018, Naveau_2019,Naveau_2019_2,Vanderhaegen_2020_Opex, Vanderhaegen_2020_OL, Vanderhaegen_2022_therm} the same scheme was used to achieve nearly optimal cancellation of the losses, thus enabling almost fully transparent propagation, here, the Raman pump power is tuned above this optimum to get a net effective gain. To calibrate the effective gain value associated to a specific Raman pump power value, we have recorded, for different multiple values of the Raman pump power, the optical time domain reflectometer (OTDR) traces obtained from the propagation of pulses. We use weak pulses to make the nonlinear effects negligible in the calibration.
Indeed, pulse powers below $50$ mW result into nonlinear lengths longer than $17$ km, which exceed the typical fiber span.

\begin{figure}[h!]
    \centering
    \includegraphics[width=\columnwidth]{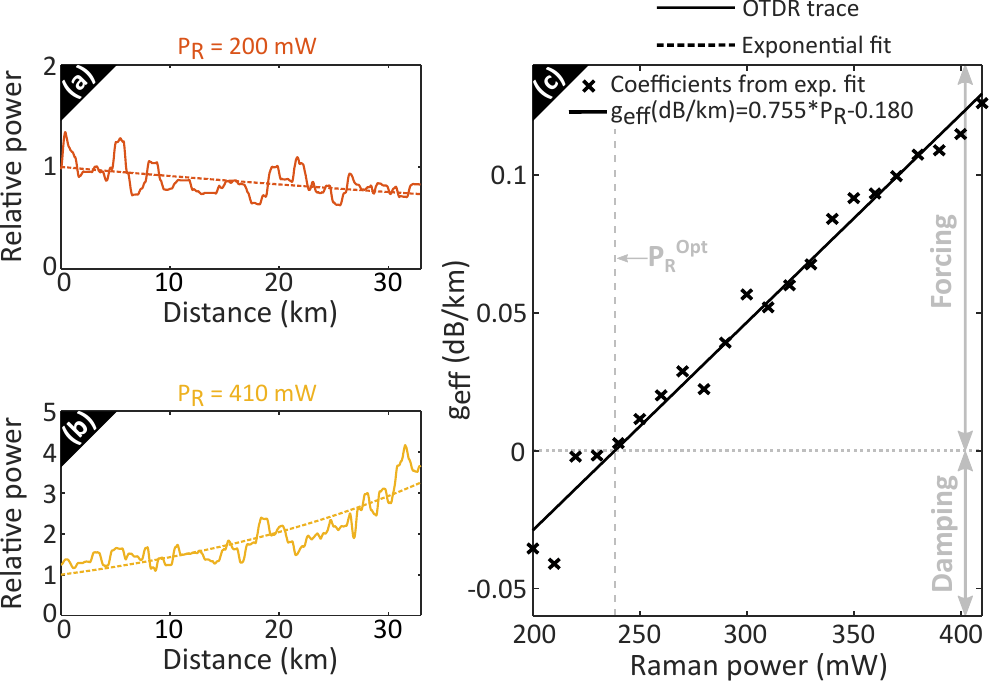}
    \caption{Relative evolutions of the backscattered signal power with (a) $P_\mathrm{R}(z=L)=200$ mW (solid orange line) and (b) $P_\mathrm{R}(z=L)=410$ mW (solid yellow line). The corresponding dashed lines stand for the exponential fit. (c) Net effective gain as a function of the Raman pump power $P_\mathrm{R}$.}
    \label{fig:fit_raman_2}
\end{figure}

Two examples of OTDR traces are plotted in Fig.~\ref{fig:fit_raman_2}(a) and (b) for an injected Raman pump power of $P_\mathrm{R}(z=L)=200$ mW and $410$ mW, respectively (these values are the Raman pump range limits). In order to evaluate the net effective gain $g_\mathrm{eff}$, we fit the observed traces by exponential curves [\textit{i.e.}, $\exp(g_\mathrm{eff} Z)$, displayed as dashed lines in Fig.~\ref{fig:fit_raman_2}(a,b)].
We find that Raman pump power below $P_\mathrm{R}^\mathrm{Opt}=240$mW results into net losses (negative slope exponential, \textit{i.e.} negative gain, see Fig.~\ref{fig:fit_raman_2}(a)), whereas above this value the exponential fit gives a net gain, as shown by dashed yellow line in Fig.~\ref{fig:fit_raman_2}(b). The results of this procedure for different values of $P_\mathrm{R}$ are summarized in Fig.~\ref{fig:fit_raman_2}(c). 
In the following observations of  separatrix crossing, we use the linear fit in Fig.~\ref{fig:fit_raman_2}(c) to assess the resulting net gain $g_\mathrm{eff}$ corresponding to the  value of $P_\mathrm{R}$ actually employed in  nonlinear experiments.


To demonstrate the influence of forcing on the nonlinear MI recurrences, we use a fully fiber-optic setup based on multi-heterodyning reflectometry, quite similar to the one used in \cite{Mussot_2018,Naveau_2019,Naveau_2019_2,Vanderhaegen_2020_OL,Vanderhaegen_2020_Opex,Vanderhaegen_2022_therm} and particularly to \cite{Vanderhaegen_2022_loss}, though with some important differences that we detail in the following. 
Since we overcompensate the losses, spontaneous MI amplification of noise may lead to thermalization \cite{Vanderhaegen_2022_therm}. In order to prevent such a detrimental effect, we use a SMF-28 fiber span of $16.73$ km (shorter compared to the span of $\sim 20$ km used in \cite{Vanderhaegen_2022_loss}) and increase the pump power to reduce the nonlinear length as well. 
This allows us to achieve both a sufficient number of recurrence cycles (at least three to be able to also observe the transition around $g_\mathrm{c2}$) within the fiber span and  a higher sideband-to-noise ratio. The signal to pump ratio is then set to $\frac{P_s(z=0)}{P_p(z=0)}=-9.5$ dB and the $P_\mathrm{R}$ maximum value to $410$ mW, which allows to get an effective gain $g_\mathrm{eff}$ up to $0.13$ dB/km. The initial relative phase we usually set to excite a shifted orbit is $\Delta \Phi_0=\pm \pi/2$ \cite{Mussot_2018}. 
However, for such an input phase, we expect from numerical simulations that $g_{c1}=0.268$ dB/km, which is beyond the achievable effective gain range. To get a smaller critical gain, we need to choose a $\Delta \Phi_0$ value closer to the separatrix, so that it requires a weaker perturbation to cross it (as predicted by the 3WM model and the finite-gap theory, see Fig.~\ref{fig:finite_gap_3wm}). We set then the initial relative phase to $\Delta \Phi_0=-0.28 \pi$. From the input pump power $P_p(z=0)=470$ mW and the fiber parameters 
$\beta_2=-21$ ps$^2$km$^{-1}$, $\gamma=1.3$ W$^{-1}$km$^{-1}$, we calculate the frequency of maximum MI gain at $f_\mathrm{peak}=38.4$ GHz. We set the modulation frequency very close to that value, at $f_m=38.2$ GHz. 

\subsection{Experimental results}
\begin{figure*}
    \centering
    \includegraphics[width=0.95\textwidth]{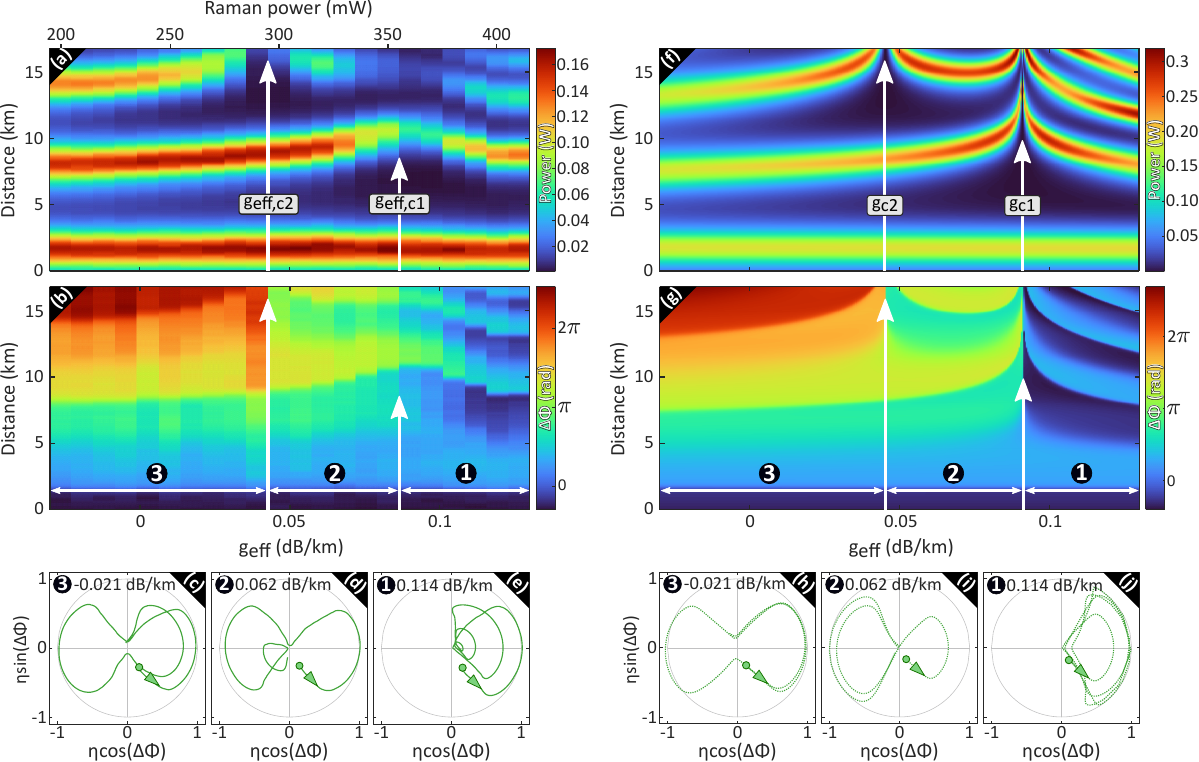}
    \caption{False color plots of (a), (f) the signal power evolution and (b), (g) the relative phase along the fiber distance as a function of the effective gain. (c-e), (h-j) Trajectories in the phase-space $(\eta \cos(\Delta \Phi),\eta \sin(\Delta \Phi))$ for $g_\mathrm{eff}=-0.021$, $0.062$ and $0.114$ dB/km. Left panels: experimental data; right panels: numerical simulations from forced NLSE Eq. (1).}
    \label{fig:exp_vs_num_gain}
\end{figure*}

Here we discuss the experimental characterization of the recurrence phenomenon in the presence of linear gain.  We display the evolution of the signal power $P_s(z)$ and the relative phase $\Delta \Phi (z)$ as a function of the effective gain in Fig.~\ref{fig:exp_vs_num_gain}(a) and (b), respectively. 
Three different regions labeled, \circled{3}, \circled{2}, \circled{1} (from left to right), are outlined. In region \circled{3}, the system is close to  optimal compensation of losses, with either a slight attenuation (down to $g_\mathrm{eff}=-0.029$ dB/km) or a slight amplification (up to $g_\mathrm{eff}=g_\mathrm{eff,c2}=0.043$ dB/km). Within the finite length of the fiber used in the experiment, in this region gain has the only  minor effect of slightly changing the recurrence distance. Looking at the power evolution in Fig.~\ref{fig:exp_vs_num_gain}(a), we observe three complete recurrences, but also a $\pi$-shift between successive conversion cycles as clearly shown by the phase evolution in Fig.~\ref{fig:exp_vs_num_gain}(b). This behavior is confirmed by the reconstructed trajectory in the phase-space in Fig.~\ref{fig:exp_vs_num_gain}(c) obtained for a specific effective gain $g_\mathrm{eff}=-0.021$ dB/km ($P_\mathrm{R}=210$ mW). The orbit oscillates from the right to the left (and viceversa) half-phase-space between two consecutive growth-decay cycles, meaning that two consecutive maximum compression points are shifted by half a temporal period. This dynamic is not surprising because it behaves exactly the same as a shifted orbit excitation in the integrable limit $g=0$. These experimental results can be compared to the numerical results in Fig.~\ref{fig:exp_vs_num_gain}(f-j). These numerical data were obtained with a slightly different initial relative phase value than the experimental one. Indeed, the numerical results with $\Delta \Phi_0=-0.28\pi$ do not predict any critical transition point for $g<0.13$ dB/km. Thus, in our numerical simulations, we  employ a slightly different value $\Delta \Phi_0=-0.335 \pi$. We suggest that such a discrepancy can simply be due to an imperfect control of the initial relative phase which is set through a Waveshaper \cite{Mussot_2018,Naveau_2021}. Indeed, while a small shift of the initial relative phase is not detrimental far from $\Delta \Phi_\mathrm{AB}$,
it can have a very strong impact on the dynamic close to it. When we account for such difference in the initial phase, we find a very good agreement between experiments and numerics in region \circled{3} (as well as in regions \circled{2}, \circled{1} described below).

By increasing the effective gain up to $g_{\mathrm{eff},c2}=0.043$ dB/km ($P_\mathrm{R}=295$ mW), we notice that the third recurrence cycle period increases so that the corresponding conversion peak does not appear within the fiber length close to $g_{\mathrm{eff},c2}$. Beyond this critical gain point, in region \circled{2}, we record three conversion peaks again. However, the phase evolution reveals that there is no longer a $\pi$ phase shift between the second and third recurrence cycles. This is confirmed by the phase-space representation in Fig.~\ref{fig:exp_vs_num_gain}(d) with $g_\mathrm{eff}=0.062$ dB/km ($P_\mathrm{R}=320$ mW). While the  trajectory initially spans the full phase-space, as expected for a P2 orbit, the third recurrence of power fraction $\eta$ occurs in the left half-plane, \textit{i.e.}, the trajectory has switched to a P1 orbit: the separatrix has been crossed between the second and third recurrence cycles. 

When approaching $g_{\mathrm{eff},c1}=0.088$ dB/km ($P_\mathrm{R}=355$ mW), the period of the second cycle also increases and the corresponding conversion peak spatially shifts at higher distance. However, beyond $g_{\mathrm{eff},c1}$, the trend inverts and the number of conversion peaks recorded within the fiber length increases up to four at $g_\mathrm{\mathrm{eff}}=0.13$ dB/km in region \circled{1}. In this region, the phase remains bounded in $[-\frac{\pi}{2}; \frac{\pi}{2}]$, similarly to what has been observed in conservative systems for a P1 orbit excitation. This is further revealed by the trajectory in the phase-space in Fig.~\ref{fig:exp_vs_num_gain}(e) for $g_\mathrm{\mathrm{eff}}=0.114$ dB/km, where all the trajectory stays in the right half-plane---all the four compression points occur in phase. In this case the separatrix has been crossed during the first growth and decay cycle, leading to an unshifted orbit excitation from the second one. The overall agreement between experiments and numerical simulations is good in regions \circled{1} and \circled{2}, except for the fact that the recorded signal power exhibit much lower power levels compared with simulations during the third and fourth recurrences (compare Figs.~\ref{fig:exp_vs_num_gain}(a) and (f)). This can be attributed to the blurring effect from slight fluctuations of the parameters but especially to noise-induced MI. Indeed, we notice a decrease of the signal power maxima, similarly to what is observed in \cite{Vanderhaegen_2022_therm}. Regarding the critical gain values, we find $g_{\mathrm{eff},c1}=0.088$ dB/km ($P_\mathrm{R}=355$ mW) and $g_{\mathrm{eff},c2}=0.043$ dB/km ($P_\mathrm{R}=295$ mW), which follow almost perfectly the inverse scaling with $n$, predicted by Eq.~\eqref{eq:finite_gap_gain}, at variance with the case of damping induced crossing \cite{Vanderhaegen_2022_loss}, which exhibited a significant discrepancy between the observed critical values of losses and their scaling with $n$. We conjecture that this could be due to the fact that, here, the system is initially excited close to A-type or P2 solutions of the NLSE, which are known to be more stable and robust to the input three-wave truncation due to their intrinsic Fourier structure \cite{Conforti_2020,Vanderhaegen_2020_OL}. 
Finally from the numerical data shown in Fig.~\ref{fig:exp_vs_num_gain}(f,g), we obtain a good agreement, the extrapolated numerical values are  $g_{c1}=0.091$ dB/km and $g_{c2}=0.045$ dB/km, though obtained for $\Delta \Phi (z=0)=-0.335 \pi$.

Finally we emphasize that weaker amplifications result into critical values of gain which become more densely spaced. However, as clear from the trend in Fig. \ref{fig:well_gain_breaking}(f), the weaker the critical gain, the longer the distance required to observe it. Therefore, in the present experiment performed at fixed fiber length, we are limited to the first two critical values of gain, since the use of longer fibers would require to improve the present Raman amplification based control of gain, an issue which will be addressed in the future.

\section{Conclusions}

Thanks to an accurate control, via counter-propagating Raman amplification, of the linear gain experienced by light during  propagation in an optical fiber, we probed the behavior of FPUT-like recurrences in the nonlinear stage of MI. We  showed  that forcing leads the shifted orbits of recurrent MI to be converted into unshifted ones. We found critical values of linear gain at which separatrix crossing occurs almost precisely after $n$ recurrence cycles: in the considered fiber span, we were able to distinguish three regions separated by the lowest-order values of $g_{cn}$, with $n=1,2$. In order to observe higher-order critical values of gain, we would need a longer fiber span and a more accurate gain profile shaping, \textit{i.e.}, a more sophisticated control of the Raman pump. Apart from the interest in taming and controlling nonlinear effects in an optical fiber, we expect to spur a fundamental interest in the behavior of other physical systems such as those with quadratic nonlinear response \cite{Trillo:23} or pendulum chains \cite{Bishop:88} that exhibit similar homoclinic/heteroclinic connection in MI.



\begin{acknowledgments}

The present research was supported by IRCICA (USR 3380 CNRS),
Agence Nationale de la Recherche (Programme Investissements d’Avenir, FARCO)
Ministry of Higher Education and Research; Hauts de France Council; European Regional Development Fund (Photonics for Society P4S, WAVETECH, FELANI, GPEG). A.A.~and S.T. acknowledge funding from the Ministry of University and Research of Italy (PRIN2020X4T57A, PRIN2022NCTCY---NextGenerationEU, with the European Union).

\end{acknowledgments}

\appendix 
\section{Exact solutions of Eq.~\eqref{eq:gcrit_3W}}
In Eq.~\eqref{eq:gcrit_3W}, we use the following formula for the unperturbed ($g=0$) evolution of the sidebands power fraction \cite{Cappellini_1991}
\begin{eqnarray} \label{solution}
\eta(z) = \frac{a_-(b_+-b_-){\rm sn}^2(\rho z \vert m) - b_-(b_+-a_-)}{(b_+-b_-){\rm sn}^2(\rho z \vert m) - (b_+-a_-)} 
\end{eqnarray}
where $\rho = \sqrt{7(a_+-b_-)(b_+-a_-)}/4$,
$m =\frac{(b_+-b_-)(a_+-a_-)}{(a_+-b_-)(b_+-a_-)}$ 
is the modulus of the Jacobian  $sn$ function
and $a_+ \ge b_+ \ge b_- \ge a_-$ 
read
$a_{\pm} = \Omega^2 \pm \sqrt{\Omega^4+4 H_0}$,
$b_{\pm} =\frac{2}{7} \left(2 -\Omega^2/2 \pm \sqrt{(2 -\Omega^2/2)^2 - 7 H_0} \right)$.


The spatial period $z_\mathrm{per}$ of the solution (\ref{solution}) can be expressed in terms of the elliptic integral of the first kind $K(m)$ as
\begin{equation} 
z_\mathrm{per} = \frac{2 K(m)}{\rho} \simeq \frac{1}{\rho} \log \frac{4}{\sqrt{1-m}},
\label{eq:3Wperiod}
\end{equation}
where the logarithmic approximation is valid for $m \simeq 1$ (\textit{i.e.}, near the separatrix). We emphasize that $z_\mathrm{per}$ is the period of power evolution while the spatial period of the figure-of-eight orbit in the phase plane (see Fig.~\ref{fig:well_gain_breaking}(a,b)) is $2z_\mathrm{per}$. In other words, $\eta(z)$ takes back its initial value twice along this type of orbits, while the relative phase $\Delta\Phi$ exhibits $\pi$-increments.

We emphasize that Eq.~\eqref{eq:gcrit_3W} gives an excellent estimate of the critical values of gain for crossing  which can be found from numerical integration of 3WM equations. However, as the full NLSE is concerned, the discrepancy discussed in Fig. 2 with the better estimate from finite-gap approach, Eq.~\eqref{eq:finite_gap_gain}, essentially arises from the deviation of the period estimated from Eq.~\eqref{eq:3Wperiod} with respect to that of the full wave dynamics, reported in \cite{Conforti_2020}.

\bibliographystyle{apsrev4-1}
\bibliography{biblio.bib}

\end{document}